\documentclass[aps,prl,reprint,showpacs,superscriptaddress]{revtex4-1}
\usepackage{graphicx}
\usepackage{color}
\usepackage{txfonts}
\usepackage{float}
\usepackage[colorlinks, linkcolor=blue, anchorcolor=blue, citecolor=blue, urlcolor=blue]{hyperref}
\usepackage{multirow}

\begin{document}


\title{Emergent 1/3 magnetization plateaus in pyroxene CoGeO$_3$}



\author{H. Guo$^{\dag}$}
\affiliation{Max-Planck-Institute for Chemical Physics of Solids, N\"{o}thnitzer Str. 40, D-01187 Dresden, Germany}
\affiliation{Neutron Science Platform, Songshan Lake Materials Laboratory, Dongguan, Guangdong 523808, China}

\author{L. Zhao$^{\dag}$}
\affiliation{Max-Planck-Institute for Chemical Physics of Solids, N\"{o}thnitzer Str. 40, D-01187 Dresden, Germany}

\author{M. Baenitz}
\affiliation{Max-Planck-Institute for Chemical Physics of Solids, N\"{o}thnitzer Str. 40, D-01187 Dresden, Germany}
\author{X. Fabr\`{e}ges}
\affiliation{Laboratoire L\'{e}on Brillouin, CEA-CNRS, CE-Saclay, 91191 Gif-sur-Yvette, France}
\author{A. Gukasov}
\affiliation{Laboratoire L\'{e}on Brillouin, CEA-CNRS, CE-Saclay, 91191 Gif-sur-Yvette, France}
\author{A. Melendez Sans}
\affiliation{Max-Planck-Institute for Chemical Physics of Solids, N\"{o}thnitzer Str. 40, D-01187 Dresden, Germany}
\author{D. I. Khomskii}
\affiliation{Physics Institute II, University of Cologne, Z\"{u}lpicher Str. 77, 50937 Cologne, Germany}
\author{L. H. Tjeng}
\affiliation{Max-Planck-Institute for Chemical Physics of Solids, N\"{o}thnitzer Str. 40, D-01187 Dresden, Germany}
\author{A. C. Komarek}
\email[]{Komarek@cpfs.mpg.de}
\affiliation{Max-Planck-Institute for Chemical Physics of Solids, N\"{o}thnitzer Str. 40, D-01187 Dresden, Germany}

\date{\today}

\begin{abstract}  
Despite the absence of an apparent triangular pattern in the crystal structure, we observe  unusually well pronounced 1/3~magnetization plateaus in the quasi one-dimensional Ising spin chain compound CoGeO$_3$ which belongs to the class of pyroxene minerals. We succeeded in uncovering the detailed microscopic spin structure of the 1/3 magnetization plateau phase by means of neutron diffraction. We observed changes of the initial antiferromagnetic zero-field spin structure that are resembling a regular formation of antiferromagnetic "domain wall boundaries", resulting in a kind of modulated magnetic structure with 1/3-integer propagation vector. The net ferromagnetic moment emerges at these "domain walls" whereas two third of all antiferromagnetic chain alignments can be still preserved. 
We propose a microscopic model on the basis of an anisotropic frustrated square lattice to explain the observations.
\end{abstract}
 
\maketitle 

Magnetization plateaus at finite magnetic field have attracted enormous attention, both theoretically and experimentally \cite{Takigawa}. The phenomenon of magnetization jumps to rational values (1/$n$) of the saturation magnetization ($M_S$) during a magnetization process is intimately connected to the presence of frustration in quantum magnets. Here one may distinguish between systems with 'geometric frustration' as a result of the crystal structure, e.g. triangular lattices, and those with 'interaction frustration' where the presence of several exchange interactions lead to a competition for the ground state \cite{Schmidt}. Examples for the first category include Dy$_2$Ti$_2$O$_7$ \cite{Sakakibara}, CdCr$_2$O$_4$ \cite{Ueda}, Ba$_3$CoSb$_2$O$_9$ \cite{Shirata}, and Ca$_3$Co$_2$O$_6$ \cite{CaCoA,CaCoB,Maignan,CaCoCC,CaCoE,CaCoH,CaCoI,CaCoJ,CaCoK,CaCoG}, and for the second category SrCu$_2$(BO$_3$)$_2$ \cite{Kageyama,Kodama,Matsuda,Corboz} and TbB$_4$ \cite{Yoshii,YoshiiB}. Other examples are listed in Refs. \cite{Takigawa,Schmidt}.

\begin{figure}
\centering
\includegraphics[width=0.8\columnwidth]{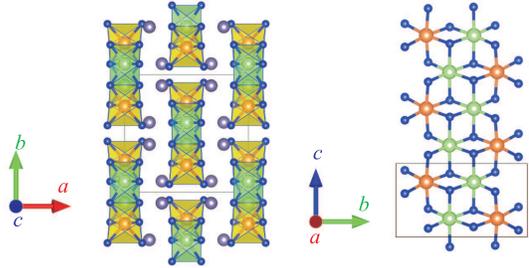}
\caption{\textbf{Crystal structure} -  Crystal structure of CoGeO$_3$ (from the \textit{c}-direction); a chain running along the \textit{c} axis is shown in the right.  Green: Co1; dark yellow: Co2; blue: oxygen; grey: Ge. The high temperature polymorph of CoGeO$_3$ has a monoclinic crystal structure with space group $C2/c$  ($a$~=~9.6623(2) \AA, $b$~=~8.9928(2) \AA, $c$~=~5.16980(10) \AA, $\beta$~=~101.2785(10)$^{\circ}$) \cite{ours}. }
\label{crystalstructure}
\end{figure}

Here, we studied the S~=~3/2 system CoGeO$_3$ which belongs to the family of pyroxene minerals \cite{Redhammer}.
Pyroxenes $A$$M$$X$$_2$O$_6$ ($A$ = mono- or divalent metal, $M$ = di- or trivalent metal, $X$ = Si$^{4+}$, Ge$^{4+}$ or V$^{5+}$) are one of the main rockforming minerals in the Earth’s crust \cite{geol,geolA,geolB,geolC} and themselves gained considerable interest due to the appearance of various interesting physical phenomena arising from their highly versatile crystal structure \cite{ooA,ooB,Jodlauk_2007}.
The pyroxene structure of monoclinic CoGeO$_3$ consist of Co$^{2+}$ ions that are forming CoO$_6$ octahedral zigzag chains running in $c$-direction - see Ref.~\cite{ours} and Fig.~\ref{crystalstructure}. These chains are located on an almost rectangular lattice in a plane perpendicular to the chains 
and there is no obvious geometric frustration in the lattice.
The system orders antiferromagnetically below T$_N$~$\sim$~33.5~K with an Ising-like character as is indicated by about $\sim$1~$\mu_B$ enhanced effective magnetic moments (compared to the expected spin-only value) and by the very anisotropic magnetic susceptibility with Weiss temperatures of opposite signs for $H$$\parallel$$c$ and $H$$\perp$$c$ \cite{ours}.

In this Letter, we report on our discovery of the emergence of extremely well-defined 1/3 magnetization plateaus in CoGeO$_3$ single crystals. 
The 1/3 value is surprising since for chains on a rectangular lattice one would rather expect a value of 1/2 \cite{Coletta}. 
This discovery enlarges the class of such plateau systems.  Moreover, we have succeeded using neutron scattering experiments on large single crystals to resolve the real magnetic structure in this 1/3 M$_S$ phase and used this information to develop a microscopic model for the formation of these 1/3 plateaus.


The floating zone growth and characterization of monoclinic CoGeO$_3$ is described in Ref.~\cite{ours}.
Magnetization measurements were performed by using the ACMS option in a physical properties measurement system (PPMS, \textit{Quantum Design})  equipped with a magnet for fields up to 9~T.
The measurements of the specific heat were carried out using a standard thermal relaxation calorimetric method in a PPMS. The dielectric properties of CoGeO$_3$ were measured on a plate ($\perp c$) with 0.5~mm thickness that was coated with silver paint on both sides. Its capacitance was measured using a high-precision bridge (AH2700A, \textit{Andeen-Hagerling Inc.}). For measuring $\varepsilon$($H$) the sample was zero-field-cooled to 10~K before starting to scan $H$ from/to $\pm$9T~ with a rate of 30~Oe/s.
Single crystal neutron diffraction measurements have been performed on the 6T2 diffractometer at Laboratoire L\'{e}on Brillouin (LLB), Saclay, France. A twined single crystal ($\sim$5 mm $\times$ 5 mm $\times$ 5 mm) was measured in fields up to 6.5~T ($H \parallel c$). The components out of the \textit{a}$^{\ast}$\textit{b}$^\ast$ (\textit{HK}) scattering plane could be reached by lifting of the detector. An area detector was used for the mapping of the $HK$ plane ($\lambda$~=~2.45~$\AA$). The integrated intensities of the zero field phase were measured by a single counter after a ZFC process, and after a FC process for the 6.5~T phase ($\lambda$~=~0.9~$\AA$). 72 nuclear and 151 magnetic reflections were measured in zero-field (ZF), and,  196 magnetic reflections were measured at 6.5~T. A nuclear structure refinement yields a volume fraction of 1:1.10(2) for the twin domains.  The \textit{Fullprof} programm package was used for magnetic structure determination.


\begin{figure}[]
\centering
\includegraphics[width=1\columnwidth]{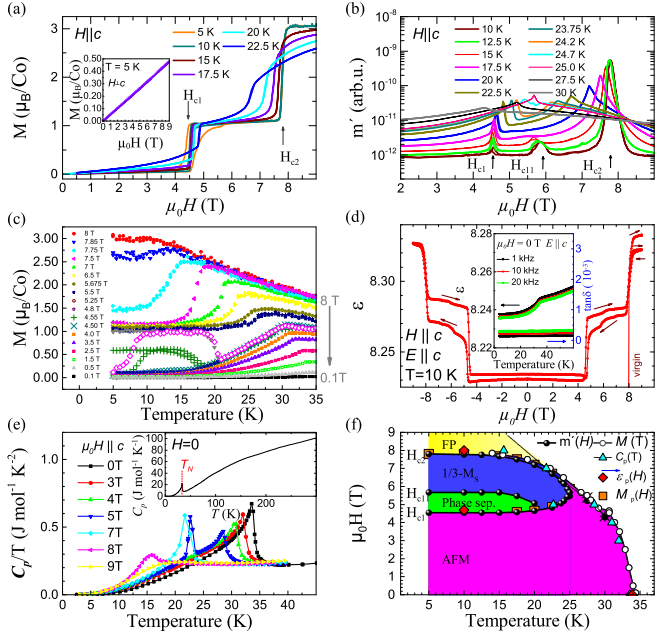}
\caption{\textbf{Phase Diagram} - (a) Magnetization $M$($H$) of CoGeO$_3$ for $H \parallel c$. (Data obtained by ramping up to 9T and down to 0T after virgin zero field cooling from 200 K down). Pronounced 1/3 magnetization plateaus can be seen. \textit{Inset:} The magnetization  for $H \perp c$ at 5~K as function of (increasing and decreasing) magnetic field which is linear in $H$ and unsaturated up to 9~T.  (b) Real part of the ac susceptibility as function of applied magnetic field (ac frequency: 100~Hz; ac field: 10~Oe).   (c)
The magnetization for various fields $H \parallel c$. The shift of the order temperature with magnetic field can be seen. At $\sim$8~T the order is completely suppressed and the system enters the field polarized (FP) regime. At intermediate fields the occurrence of the 1/3 magnetization plateau becomes visible. (d) Field dependence of $\varepsilon$ at T~=~10~K ($\mu_0 H$ was scanned in the range of -9~T and 9~T with a rate of 30 Oe/sec.) \textit{Inset:} Temperature dependence of the dielectric constant $\varepsilon$ and the corresponding dielectric loss tan$\delta$ in zero field.  (e) Specific heat $C_p$(T) of CoGeO$_3$ measured for different external fields ($H$~= 0~T to 9~T) along c-axis. (e) Magnetic phase diagram of CoGeO$_3$ for $H \parallel c$. (Lines are guide to the eyes.)}
\label{phase}
\end{figure}


Direction dependent magnetization measurements on our CoGeO$_3$ single crystals at low temperatures reveal
a sharp step-like increase of the magnetization at  $H_{c1}$~$\sim$4.5~T for fields parallel to the $c$-axis - see Fig.~\ref{phase}(a);  
(for fields $H$ perpendicular to the $c$-axis a linear $M$-$H$ relationship can be observed as is shown in the inset).
After this first jump of the magnetization $M$ a value of roughly 1~$\mu_B$ per Co$^{2+}$ ion is attained and a very well pronounced magnetization plateau appears up to the next critical field $H_{c2}$~$\sim$8~T where the system becomes fully field polarized (FP) with $M$ $\sim$ 3~$\mu_B$ per Co$^{2+}$ ion.
These observations resemble the ones in
Ca$_3$Co$_2$O$_6$, \cite{CaCoA,CaCoB,CaCoC,CaCoD,CaCoE} and CoV$_2$O$_6$ \cite{CoVaA,CoVaB,CoVaC,CoVaD,CoVaE,CoVaF,CoVaG}. However, the detailed underlying magnetic structure was not directly measured in the magnetization plateau (1/3 M$_S$) phase although the zero-field AFM phase was determined accurately. 

We established the entire magnetic phase diagram of CoGeO$_3$ in great detail with dc magnetization, ac susceptibily, specific heat and dielectric measurements - see Fig.~\ref{phase}(a-f). Besides the antiferromagnetic (AFM) phase, the 1/3 M$_S$ phase and the FP phase we observe yet another phase that we assign to magnetic phase separation, see Fig.~\ref{phase}(f).

\begin{figure}[!b]
\centering 
\includegraphics[width=0.7\columnwidth]{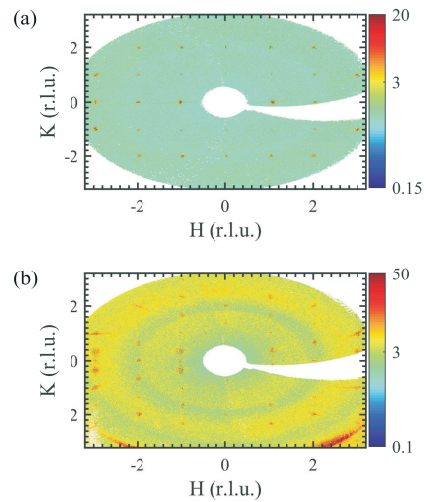}
\caption{\textbf{Neutron Maps} -  Neutron scattering intensity maps within the $HK0$ scattering plane measured at (a) 0 T and (b) 6.5 T using an area detector. }  
\label{map}
\end{figure}

\begin{figure}[!b]
\centering 
\includegraphics[width=1\columnwidth]{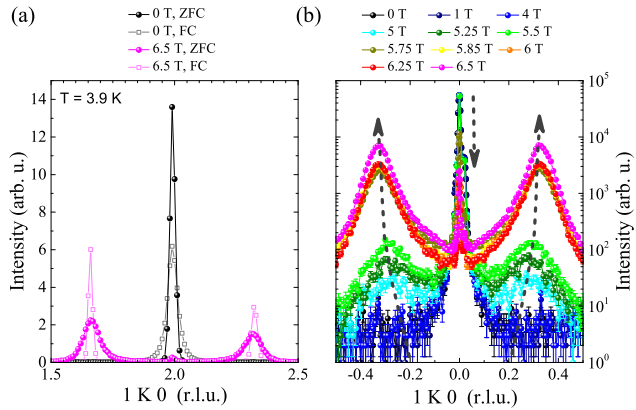}
\caption{\textbf{Neutron Scans} -  (a) Scans across the magnetic peak positions in $K$-direction. (b) A detailed magnetic field dependence (within a ZFC process). } 
\label{scans}
\end{figure}

Furthermore, we performed neutron scattering experiments - first at 3~K after a zero field cooling (ZFC) process, see Fig.~\ref{map}(a).  
In zero-field (ZF) superlattice reflections with odd values of $H + K$  appear. This indicates a breaking of the C-centering due to the magnetic ordering. Superlattice reflections with $|L|>0.05$ were absent. A magnetic propagation vector \textbf{k}$_\mathrm{ZF}$ = \textbf{k}$_1$ = (1~0~0) is in agreement with these observations and also with powder diffraction data \cite{Redhammer}.
Then we applied an external magnetic field of 6.5~T along the crystallographic $c$-axis and observed a magnetic peak splitting in $K$-direction when entering the regime of the 1/3 M$_S$ phase. As a consequence the magnetic propagation vector in the 1/3 M$_S$ phase changes to  \textbf{k}$_\mathrm{6.5 T}$ = \textbf{k}$_2$ = (1~1/3~0) - see Fig.~\ref{map}(b).
Scans along the $K$-direction are shown in Fig.~\ref{scans}(a).
As can be seen in Fig.~\ref{scans}(a), the magnetic peaks at third-integer positions are distinctly broader after an initial ZFC process than after a field cooling (FC) process (and vice versa for the integer peak). In Fig.~\ref{scans}(b) a more detailed field dependence is shown. During a ZFC process magnetic phase separation appears in CoGeO$_3$ with two magnetic phases phases - with \textbf{k}$_1$ and \textbf{k}$_2$ - appearing simultaneously. The peak widths of roughly $\sim$0.07~r.l.u. at 6.5~T  indicate short range magnetic correlations and, hence, magnetic (nano) phase separation - with coexistence of AFM (majority) phase and 1/3~M$_S$ (minority) phase.

\begin{figure}[!t]
\centering
\includegraphics[width=1\columnwidth]{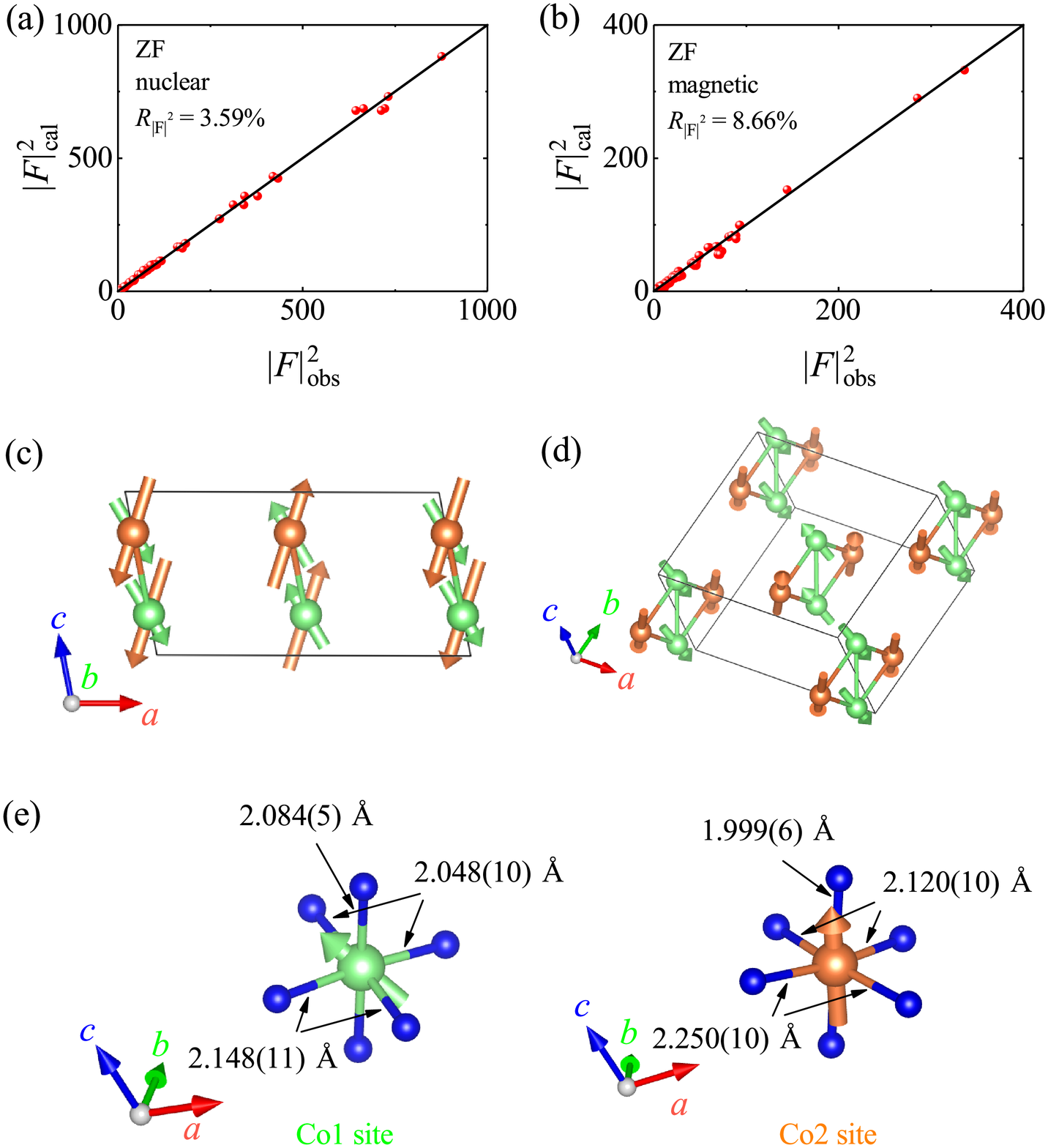}
\caption{\textbf{Neutron refinement results for the zero field phase} - Calculated structure factor $|F|_{cal}^2$ against the observed structure factor $|F|_{obs}^2$ for (a) the nuclear structure and (b) the magnetic structure refinement at 0~T. The corresponding magnetic structure is shown in (c) in a view from the $b$ direction, and in (d) for different perspective. (e) Local coordination of the two Co sites; green: Co1; dark yellow: Co2; blue: oxygen. \label{refinementA}}

\end{figure}

\begin{figure}[!t]
\centering
\includegraphics[width=0.88\columnwidth]{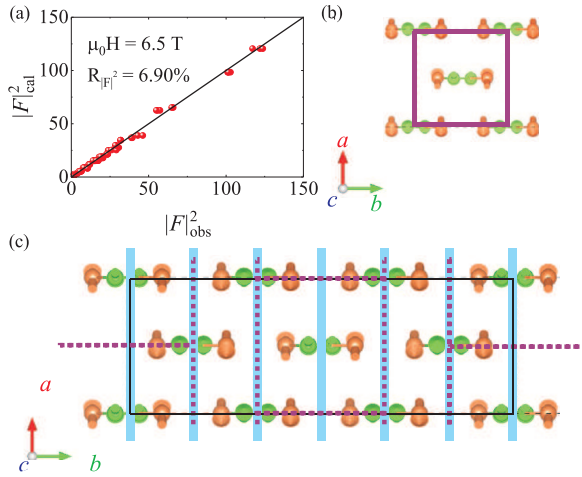}
\caption{\textbf{Neutron refinement results for the 1/3 magnetization plateau phase} -  Calculated structure factor $|F|_{cal}^2$ against the observed structure factor $|F|_{obs}^2$ for the magnetic structure refinement at 6.5~T. (b) Zero-field magnetic structure. The purple lines indicate the AFM unit cell. (c) The magnetic structure at 6.5~T with a $a \times 3b \times c$ magnetic unit cell (black lines). The light blue lines indicate planes of ferromagnetically aligned Co chains. The dashed purple lines indciate regions where the spins are aligned like in the zero-field magnetic unit cell.  \label{refinementB}}

\end{figure}

\begin{table}[!h] 
\caption{IR with basis vectors $\psi_\nu$ for \textbf{k}~=~(1 0 0); Co ions at the $4e$ sites: Co1\_1(0~$y_1$~0.75), Co1\_2(0~1-$y_1$~0.25) and Co2\_1(0~$y_2$~0.75), Co2\_2(0~1-$y_2$~0.25) with $y_1$ = 0.0922 and $y_2$ = 0.2703.\label{BV}}
 
\begin{tabular}{lcccccc}
IRs &  $\psi_\nu$  & {Co$i$\_1  }  & {Co$i$\_2  } &$\psi_\nu$ & {Co$i$\_1  }  & {Co$i$\_2  } \\
 \hline
{$\Gamma_1$ } &  {$\psi_1$ }  & {(0~1~0) } & {(0~1~0) }  & & &  \\
{$\Gamma_2$ } &  {$\psi_2$ }  & {(0~1~0) } & {(0~-1~0) } \\ 
{$\Gamma_3$ } &  {$\psi_3$ }  & {(1~0~0) } & {(1~0~0) }  &    {$\psi_4$ }  & {(0~0~1) } & {(0~0~1) }  \\
{$\Gamma_4$ } &  {$\psi_5$ }  & {(1~0~0) } & {(-1~0~0) } &   {$\psi_6$}  & {(0~0~1) } & {(0~0~-1) } \\
\end{tabular}

\end{table}

\begin{table}
  \caption{Refinement results of the magnetic structure for CoGeO$_3$ at zero field and at 6.5 T; values in units of $\mu_B$. \label{value}}
  \begin{tabular}{rlllcrlll}
    \multicolumn{2}{c}{}                     &    0 T       &  6.5 T     &   &   &    &    0 T       &  6.5 T   \\
  \hline
  \multirow{3}{*}{ } &  $M_a$         &   0.98(7)    &  0.98(7)  & &    &  $M_a$         &  -2.13(7)    &  2.09(8)  \\
                     Co1\_1       &  $M_c$         &  -2.51(3)    &  2.49(4)  &  &  Co2\_1  &  $M_c$         &  -4.27(3)    &  4.12(3)  \\
                            &  $M_{tot}$     &   2.86       &  2.85     &   &  &  $M_{tot}$     &   4.39       &  4.24    \\
  \hline
\end{tabular}
\end{table} 

Finally, we analyzed the spin structure in the ZF AFM phase and in the 1/3 M$_S$ phase.
For space group $C2/c$ and propagation vector \textbf{k}$_1$ = (1~0~0), the magnetic representation for the Co ions at the $4e$ site decomposed into four one dimensional irreducible representations as $\Gamma = \Gamma_1 + \Gamma_2 + 2\Gamma_3 + 2\Gamma_4$, see Tab.~\ref{BV} for the corresponding basis vectors. For a second order transition an ordering of the two Co ions with same IR is expected. Finally, the AFM structure can be described by $\Gamma_3$ with moments aligned within the $ac$ plane. Plots of $|F_\mathrm{cal}|^2$ vs. $|F_\mathrm{obs}|^2$ are shown in Fig.~\ref{refinementA}(a,b) and the observed spin structure is shown in Fig.~\ref{refinementA}(c,d); magnetic moments are listed in Tab.~\ref{value}.
The Co1 and Co2 ions are aligned ferromagnetically along $c$ but antiferromagnetically along $a$. The moments of the Co2 ions point along the shortest Co-O bond direction.
The Co1 moments attain their spin only values whereas an additional orbital contribution has to be considered for the Co2 moments \cite{newA,newB,newC}.

The spin structure of the 1/3 M$_S$ phase ($a \times 3b \times c$) could be determined as well - see Fig.~\ref{refinementB}.
Compared to the AFM structure, the moments in
entire Co-chains are flipped in a way that ferromagnetically ordered chains are aligned within planes that are running in $ac$-direction and with inverted magnetic moments in every third of these planes with their sizes remaining basically the same as for the zero field phase - see Tab.~\ref{value}.
The total value of the net magnetization at 6.5~T amounts to $M_c$~=~$(2.49 + 2.49 - 2.49 + 4.12 + 4.12 - 4.12 )/6$~$\mu_B$/Co~$=$~1.1~$\mu_B$/Co which is consistent with the observation of 1/3 magnetization plateaus.


Our study unravels the nature of a 1/3 magnetization plateau phase which emerges from the zero field AFM structure by flipping half of all sheets of ferromagnetic chains in order to minimize the energy  at high magnetic fields.
Although half of all chains are flipped, only one third of all of the antiferromagnetic alignments of nearest neighboring Co chains of the ZF magnetic structure are lost. This is realized by the simultaneous flipping of three neighboring layers of
Co chains. As one can see from the top-view in Fig.~\ref{refinementB}(c) there exist always three neighboring layers of Co spins (light blue lines) that still show the ZF AFM ordering pattern within the 1/3 M$_S$ phase. This situation resembles the creation of antiferromagnetic domains  which are  in one direction of the size of one unit cell. In this picture the net ferromagnetic moment appears at the "domain wall boundaries".

\begin{figure}
\centering
\includegraphics[width=0.9\columnwidth]{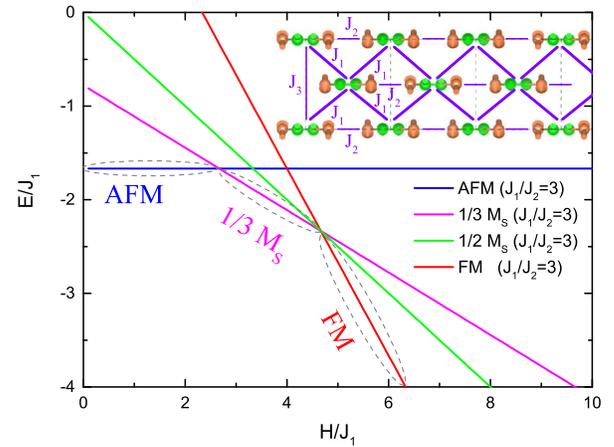}
\caption{\textbf{frustrated square lattice} - Calculated energies for an anisotropic frustrated square lattice with one diagonal exchange interaction. The 1/2~M$_\mathrm{S}$ phase is never stable. For a ratio of the values of AFM nn- and nnn-exchange interactions around 3 the expected phase diagram resembles the one shown in Fig.~\ref{phase}(f). The inset visualizes the underlying model of a frustrated square lattice with diagonal exchange. \label{theor}}

\end{figure}

If one plane more or one plane less would be flipped the net moment would be zero whereas the amount of AFM alignments would stay the same.
Therefore, the 1/3 $M_S$ phase can be lower in energy compared to such other antiferromagnetic phases.
Moreover, it could explain why there is not yet another intermediate magnetization plateau phase.
At the verge of antiferromagnetism and 1/3 $M_S$ phase magnetic (nano) phase separation appears which shows that
other magnetic structures where more than three layers are flipped at a time are not energetically
favorable compared to magnetic phase separation with AFM phase and 1/3 $M_S$ phase.

This energetic stability of the 1/3 $M_S$ phase can be further supported by a theoretical model where one treats each predominantly ferromagnetic Co chain \cite{PhysRevB.77.064405,Streltsov} running in $c$-direction as a single classical spin. In the $ab$-plane these spins form an effective anisotropic square (in general rhombic) lattice - see the inset in Fig.~\ref{theor} - with strong nearest neighbor (nn) exchange interaction $J_1$ and one weaker diagonal next nearest neighbor (nnn') exchange interaction $J_2$. The other diagonal exchange interaction $J_3$ shall be that weak $J_3 \ll J_2$ that we can neglect it. 
This model is justified because CoGeO$_3$ consist of an Co1 zig-zag chain with further Co2 ions attached to these chains. Thus, the distances between two chains are quite short in one diagonal direction $\parallel$$b$ ($J_2$) but much larger in the other diagonal direction $\parallel$$a$ ($J_3$) - see Fig.~\ref{theor}.
For this model with Ising spins (as is typical for Co$^{2+}$ ions), $H$~$=$~$J_1 \sum_{nn}^{}$${{S_i} {S_j}}$~$+$~$J_2 \sum_{nnn'}^{}$${{S_i} {S_j}}$~$-$~$H S$, one can calculate the energies for the AFM phase, for the 1/3~M$_\mathrm{S}$ phase that we determined by means of neutron diffraction, for the field polarized (FM) phase and also for a theoretical 1/2~M$_\mathrm{S}$ phase.
The energies for these different phases are plotted in Fig.~\ref{theor}.
Obviously, a 1/2~M$_\mathrm{S}$ phase is not stable for any external field $H$.
Instead, a phase diagram that is resembling our observations in Fig.~\ref{phase}(f) can be expected for this frustrated (anisotropic) square lattice with one diagonal exchange. Without the need for detailed \textit{ab-initio} calculations our simple model calculations are already sufficient to clarify the origin of the 1/3 magnetization plateau phase.

Concluding, we observed very well pronounced 1/3 magnetization plateau in the synthetic pyroxene material CoGeO$_3$.
The nature of the 1/3~M$_\mathrm{S}$ plateau phase could be unravelled by means of single crystal neutron diffraction.
The minimization of energy in an external field $H$$\parallel$$c$ (the magnetic easy axis direction) is realized by a modulated magnetic structure with 1/3-integer magnetic propagation vector that resembles a regular formation of "domain walls" within the original N\'{e}el structure of the effective Ising square lattice. A model based on a frustrated anisotropic square lattice with one diagonal exchange reveals the geometrically frustrated nature of this system.

\section{acknowledgement}
We thank P. Thalmeier for fruitful discussions.
The research in Dresden was partially supported by the Deutsche Forschungsgemeinschaft through SFB 1143 (Project-Id 247310070). The work of D. Kh. was funded by the Deutsche Forschungsgemeinschaft (DFG, German Research Foundation) - Project number 277146847 - CRC 1238.
\bibliography{CoGeO}

\begin{thebibliography}{45}
\bibitem [{new()}]{newC}%
  \BibitemOpen
  \href@noop {} {\bibinfo  {journal} {The local coordination of the Co2 ion can
  be approximated as a compressed octahedron of oxygens. Defining the local
  $x$, $y$, $z$ axes along the Co-O bonds and the compression axis as the
  $z$-axis, the high spin Co$^{2+}$ 3d$^7$ configuration will have (in addition
  to the five majority-spin orbitals) the minority spin $xy$ occupied due to
  the lower crystal field energy. The seventh orbital to be occupied is then
  given by the linear combination of the $yz$ and $zx$ which forms the $d_1$
  state and thus carries a large orbital moment [41,42]. The local coordination
  of the Co1 ion is more complex and apparently, the crystal field is such that
  the orbital degeneracy in the $t_{2g}$ subshell is fully lifted with energy
  separations that are sufficiently large in comparison with the spin-orbit
  interaction}\ }\BibitemShut {NoStop}%
\end{thebibliography}
\end{document}